\documentclass[twocolumn]{aastex63}
\usepackage{multirow}
\usepackage[figuresright]{rotating}
\usepackage{subfigure}
\usepackage{epic,eepic}
\usepackage{graphicx}
\usepackage{longtable}
\usepackage{float}
\usepackage{lineno}
\usepackage{pifont}
\usepackage{gensymb}
\usepackage{multirow}
\usepackage{amsmath}
\usepackage{color}

\shorttitle{}
\shortauthors{}

\begin{document}

\title{Stellar cycle and evolution of polar spots in an M+WD binary}

\correspondingauthor{Song Wang}
\email{songw@bao.ac.cn}

\author{Xinlin Zhao}
\affiliation{Key Laboratory of Optical Astronomy, National Astronomical Observatories, Chinese Academy of Sciences, Beijing 100101, China}
\affiliation{School of Astronomy and Space Sciences, University of Chinese Academy of Sciences, Beijing 100049, China}

\author{Song Wang}
\affiliation{Key Laboratory of Optical Astronomy, National Astronomical Observatories, Chinese Academy of Sciences, Beijing 100101, China}
\affiliation{Institute for Frontiers in Astronomy and Astrophysics, Beijing Normal University, Beijing 102206, China}

\author{Xue Li}
\affiliation{Key Laboratory of Optical Astronomy, National Astronomical Observatories, Chinese Academy of Sciences, Beijing 100101, China}
\affiliation{School of Astronomy and Space Sciences, University of Chinese Academy of Sciences, Beijing 100049, China}

\author{Yue Xiang}
\affiliation{Yunnan Observatories, Chinese Academy of Sciences, Kunming 650216, PR China}
\affiliation{Key Laboratory for the Structure and Evolution of Celestial Objects, Chinese Academy of Sciences, Kunming 650216, China}

\author{Fukun Xu}
\affiliation{Yunnan Observatories, Chinese Academy of Sciences, Kunming 650216, PR China}
\affiliation{Key Laboratory for the Structure and Evolution of Celestial Objects, Chinese Academy of Sciences, Kunming 650216, China}

\author{Shenghong Gu}
\affiliation{Yunnan Observatories, Chinese Academy of Sciences, Kunming 650216, PR China}
\affiliation{Key Laboratory for the Structure and Evolution of Celestial Objects, Chinese Academy of Sciences, Kunming 650216, China}
\affiliation{School of Astronomy and Space Sciences, University of Chinese Academy of Sciences, Beijing 100049, China}

\author{Jifeng Liu}
\affiliation{Key Laboratory of Optical Astronomy, National Astronomical Observatories, Chinese Academy of Sciences, Beijing 100101, China}
\affiliation{School of Astronomy and Space Sciences, University of Chinese Academy of Sciences, Beijing 100049, China}
\affiliation{Institute for Frontiers in Astronomy and Astrophysics, Beijing Normal University, Beijing 102206, China}
\affiliation{WHU-NAOC Joint Center for Astronomy, Wuhan University, Wuhan, Hubei 430072, China}

\begin{abstract}

Stellar activity cycles reveal continuous relaxation and induction of magnetic fields.
The activity cycle is typically traced through the observation of cyclic variations in total brightness or Ca H\&K emission flux of stars, as well as cyclic variations of orbital periods of binary systems.
In this work, we report the identification of a semi-detached binary system (TIC 16320250) consisting of a white dwarf (0.67 $M_{\odot}$) and an active M dwarf (0.56 $M_{\odot}$). 
The long-term multi-band optical light curves spanning twenty years revealed three repeated patterns, suggestive of a possible activity cycle of about ten years of the M dwarf.
Light curve fitting indicates the repeated variation is caused by the evolution, particularly the motion, of polar spots.
The significant Ca H\&K, H$\alpha$, ultra-violet, and X-ray emissions imply that the M dwarf is one of the most magnetically active stars.
We propose that in the era of large time-domain photometric sky surveys (e.g., ASAS-SN, ZTF, LSST, Sitian), long-term light curve modeling can be a valuable tool for tracing and revealing stellar activity cycle, especially for stars in binary systems.

\end{abstract}


\section{INTRODUCTION}
\label{intro.sec}

The stellar cycle can be traced by chromospheric activity variation \citep{1995ApJ...438..269B}, orbital period variation (i.e., $O-C$ diagram) \citep{1994PASP..106.1075R}, or polarity switch from Doppler imaging or Zeeman Doppler imaging \citep{2011AN....332..866M}.
On the one hand, similar to our sun, stellar activity cycles are generally multi-periodic \citep{2009A&A...501..703O}.
On the other hand, different methods may lead to different cycle estimations.
As an example, the cycle lengths unveiled by direct tracking of polarity switches are sometimes significantly shorter than those derived from chromospheric activity monitoring \citep{2011AN....332..866M}.
Recently, the cycles discovered by CoRoT \citep{2015A&A...583A.134F} or Kepler \citep{2014MNRAS.441.2744V} missions are much shorter ($\approx$ 2--3 years) than classical activity cycle lengths.
Although the mechanism driving the activity cycle is still unknown, the existence of multiple cycles in one star suggests different underlying dynamos can operate simultaneously \citep{2017ApJ...845...79B}.

During a stellar cycle, the number and location of spots on the surface of stars may change due to the variation in the magnetic field geometry.
The famous butterfly diagram, described by the solar spots, reflects the existence of the 11-year solar cycle.
However, for distant stars, some spots undergo significant variation over the stellar cycle \citep{2013A&A...553A..40H,2003AN....324..202R}, while others do not \citep{2001ASPC..223..895A,1994A&A...281..811I}.
In general, light curve fitting \citep{1992A&A...259..183S} and Doppler mapping \citep{1987ApJ...321..496V} using high-resolution spectra are often used to derive the parameters of stellar spots.

We proposed to measure the stellar cycle by searching for repeated patterns in long-term photometric light curves of binary systems (containing one unseen compact object), which is caused by the motion and appearance (or disappearance) of spots.
Compared with single stars, the orbital motion of the binary (e.g., inferior conjunction, quadrature, and superior conjunction), like anchors, can be utilized to accurately position the stellar spot and verify its stability over a long time.
Furthermore, the multi-band ellipsoidal light curves of a binary system can help determine the inclination angle of the orbital plane, which is generally coplanar with the stellar rotation plane.
Therefore, the visible star's spot properties can be measured with a higher degree of confidence compared to single stars.

In this paper, we identified a semi-detached binary TIC 16320250 (R.A. = 231.951995 deg; Dec. = +35.615920 deg) containing an active M dwarf and a white dwarf companion.
The long-term light curves reveal repeated patterns caused by the evolution of polar spots on the surface of this M dwarf.
This paper is organized as follows.
Section \ref{star.sec} describes the spectral observations and the properties of the M dwarf.
Section \ref{lc_sc.sec} introduces the long-term light curves and the stellar cycle of this system.
In Section \ref{orbit.sec}, we derived the parameters of the Kepler orbit and stellar spots by radial velocity fitting and light curve fitting.
Section \ref{discuss.sec} discusses the evolution of polar spots on the M dwarf, the nature of the companion and the stellar activity of this M dwarf.
Finally, We present a summary of our results in Section \ref{summary.sec}.

\section{Spectroscopic observations and stellar parameters}
\label{star.sec}

\subsection{Spectral observation}

From Mar 9, 2015 to Mar 21, 2021 we obtained 8 low-resolution spectra (LRS; $R \sim 1800$) and 10 medium-resolution (MRS; $R \sim 7500$) spectra of TIC 16320250 using LAMOST.
The raw CCD data were reduced by the LAMOST 2D pipeline, including bias and dark subtraction, flat field correction, spectrum extraction, sky background subtraction, wavelength calibration, etc \citep{2015RAA....15.1095L}.
The wavelength calibration of the data was based on the Sr and ThAr lamps and night sky lines \citep{2010ApJ...718.1378M}.
The reduced spectra used the vacuum wavelength scale and had been corrected to the heliocentric frame.
We carried out seventeen observations using the Beijing Faint Object Spectrograph and Camera (BFOSC) mounted on the 2.16 m telescope at the Xinglong Observatory. 
The observed spectra were reduced using the IRAF v2.16 software \citep{1986SPIE..627..733T,1993ASPC...52..173T} following standard steps, and the reduced spectra were then corrected to vacuum wavelength.

\begin{figure}[htbp!]                                                     
\center 
\includegraphics[width=0.48\textwidth]{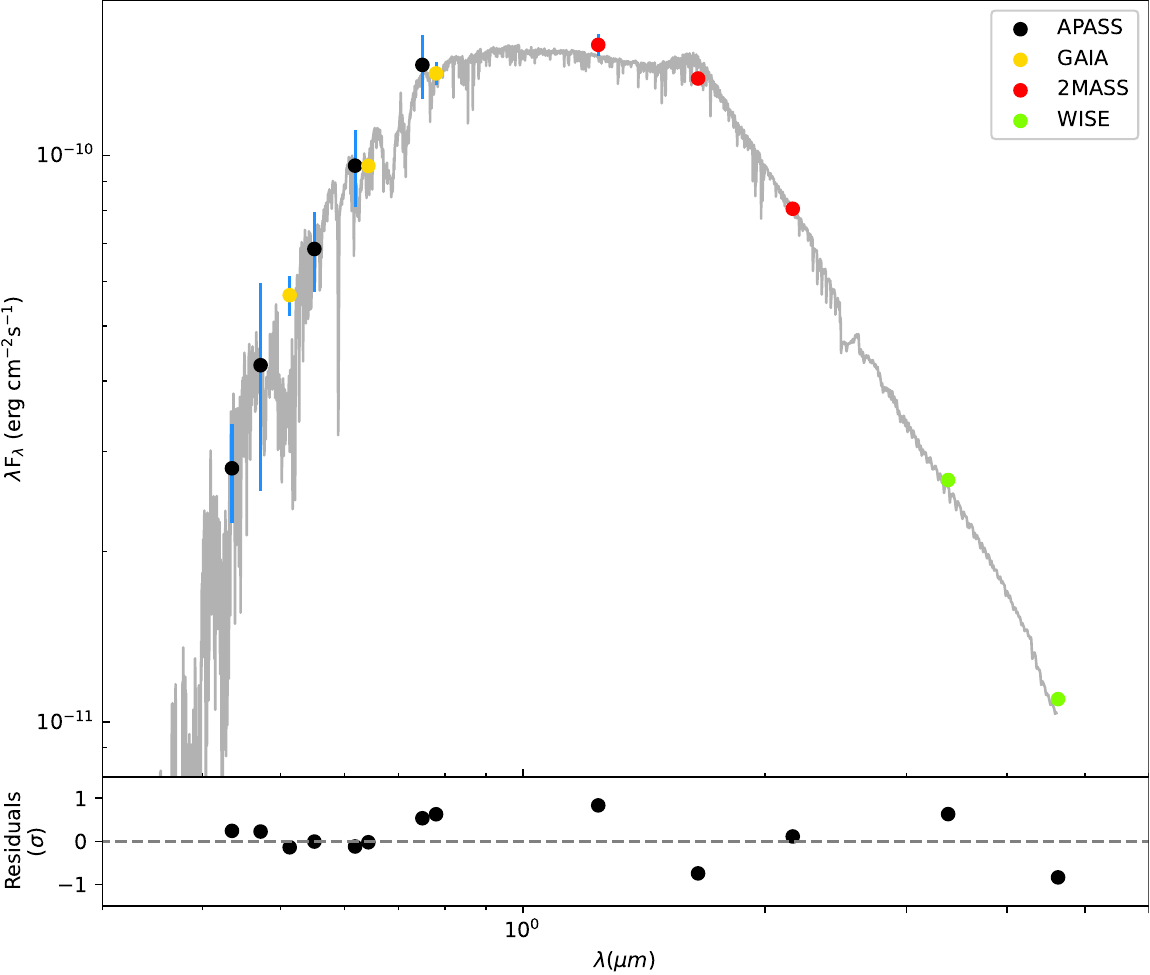}
\caption{SED fitting of TIC 16320250.  The observed data come from APASS (black dots), {\it Gaia} (yellow dots), 2MASS (red dots) and WISE (green dots). The gray line is the best model.}
\label{sed.fig}
\end{figure}

\subsection{Stellar parameters}
\label{para.sec}

The {\it Gaia} EDR3 gives a parallax of $\varpi = 8.4484 \pm 0.0111$ mas \citep{2021A&A...649A...1G}, corresponding to a distance of 118.01$\pm0.16$ pc \citep{2021AJ....161..147B}.
The $E(B-V)$ value is nearly zero, calculated with $E(B-V) = 0.884 \times {\rm (Bayestar19)}$, the latter\footnote{http://argonaut.skymaps.info/usage} of which ($\approx$0.00$\pm$0.01) is derived from the Pan-STARRS DR1 dust map \citep{2015ApJ...810...25G}.

Both the DD-Payne method \citep{2019ApJS..245...34X} and the Stellar LAbel Machine (SLAM) method \citep{2021ApJS..253...45L} have determined stellar parameters using LAMOST low-resolution spectra.
For TIC 16320250, the parameters derived by DD-Payne are $T_{\rm eff} = 4296\pm25$ K, log$g$ $= 4.39\pm0.04$ and [Fe/H] $= -0.53\pm0.03$, while the parameters derived by SLAM are $T_{\rm eff} =$ 4241$\pm$36 K and [M/H] $=$ $-0.60 \pm$0.09. 
The LAMOST DR9 presents an estimation of the atmospheric parameters from one medium-resolution observation with the LASP pipeline, as $T_{\rm eff} =$ 4019 K, log$g$ $=$ 4.76 and [Fe/H] $= 0.03$.
For cool stars, the [Fe/H] estimations from the DD-Payne and SLAM methods are smaller than the values from the LASP method, mainly due to different training sets \citep{2021RAA....21..292W}.

We used two other methods to further constrain the atmospheric parameters.
First, we tried the {\it astroARIADNE} python module which performs spectral energy distribution (SED) fitting.
Multi-band magnitudes, including APASS ($B$, $V$, $g$, $r$, $i$), {\it Gaia} ($BP$, $G$, $RP$), 2MASS ($J$, $H$, $K_{\rm S}$) and $WISE$ ($W$1, $W$2), together with the {\it Gaia} parallax and foreground extinction (set as 0.01), were used in the SED fitting.
We applied five atmospheric models
(PHOENIX\footnote{https://phoenix.astro.physik.uni-goettingen.de/}, 
BT-Settl\footnote{http://osubdd.ens-lyon.fr/phoenix/Grids/}, 
Kurucz\footnote{http://ssb.stsci.edu/cdbs/tarfiles/synphot4.tar.gz}, 
CK04\footnote{http://ssb.stsci.edu/cdbs/tarfiles/synphot3.tar.gz}, 
Coelho\footnote{http://specmodels.iag.usp.br/}) 
and the BT-Settl model returned the best-fit results (Figure \ref{sed.fig}).
The derived atmospheric parameters were given as 
$T_{\rm eff} = 3954^{+46}_{-45}$ K, log$g$ $=$ $4.66^{+0.06}_{-0.03}$ and [Fe/H] $= -0.14^{+0.18}_{-0.16}$. 
Although the M star has filled its Roche lobe (Section \ref{lcfit.sec}), the disk around the companion star shows no clear contribution to the SED, which means the atmospheric parameters derived from SED fitting are reasonable.

Second, we performed the {\it isochrones} Python module \citep{2015ascl.soft03010M} which fits the photometric or spectroscopic parameters with MIST models and returns observed and physical parameters. The input priors include the effective temperature, surface gravity, metallicity, multi-band magnitudes ($Gaia$ and 2MASS), {\it Gaia} parallax and extinction $A_V$ (set as 0.01). The derived parameters are 
$T_{\rm eff} = 3974^{+147}_{-143}$ K, log$g$ $=$ $4.72^{+0.06}_{-0.04}$ and [Fe/H] $= -0.01^{+0.15}_{-0.16}$. 

We finally averaged the parameters from above estimations (DD-Payne, LASP, {\it astroARIADNE}, and {\it isochrones}):
$T_{\rm eff} = 4061\pm138$ K, log$g$ $=$ $4.63\pm0.14$ and [Fe/H] $= -0.16\pm0.22$.

\begin{figure*}[ht!]  
\center 
\includegraphics[width=0.98\textwidth]{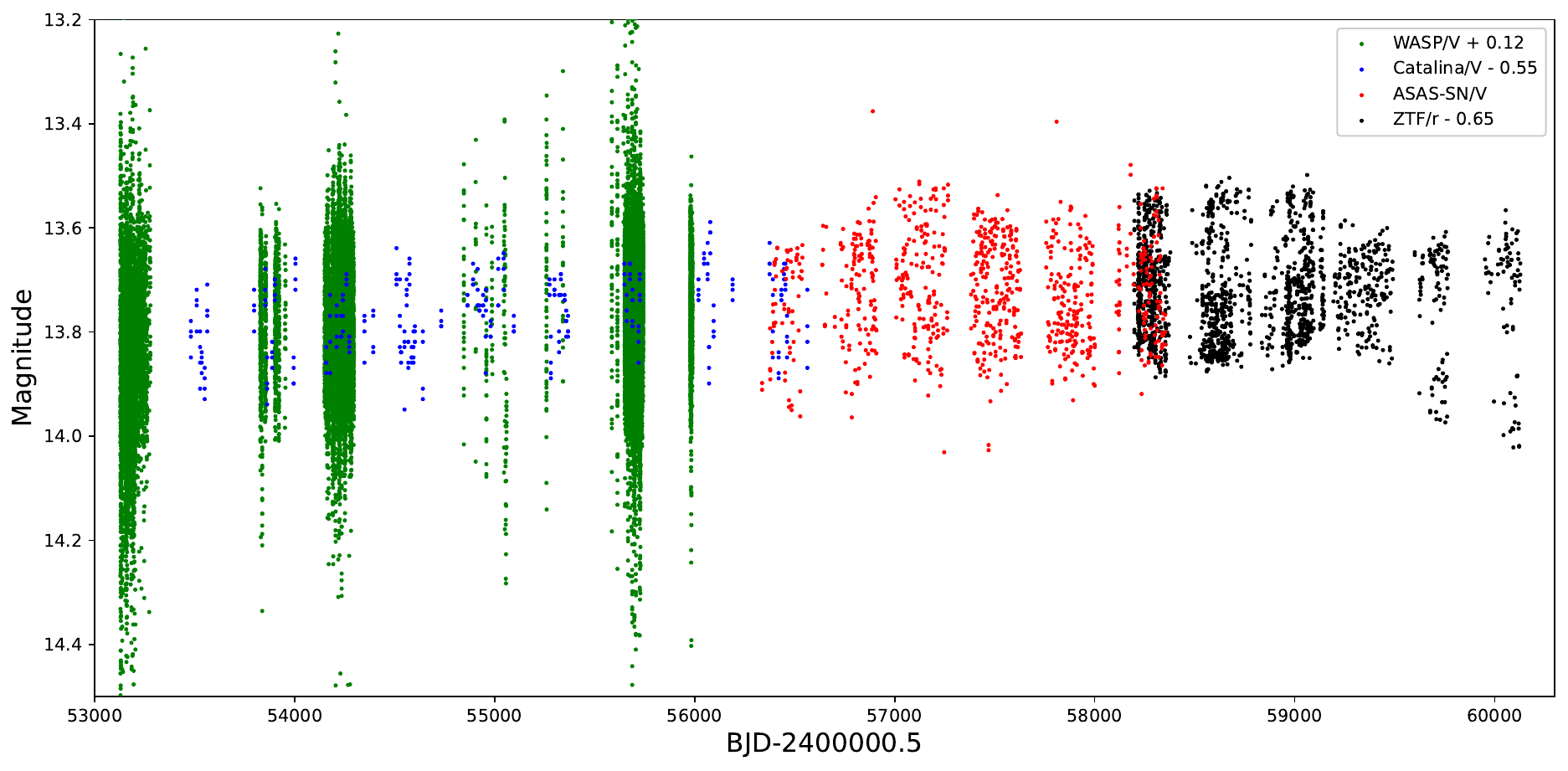}
\caption{Long-term light curves of TIC 16320250. The $V$ magnitudes are from superWASP (green dots), Catalina (blue dots), and ASAS-SN (red dots). The $r$ magnitudes are from ZTF (black dots).}
\label{Vobs.fig}
\end{figure*}

\begin{figure*}
\center 
\includegraphics[width=0.98\textwidth]{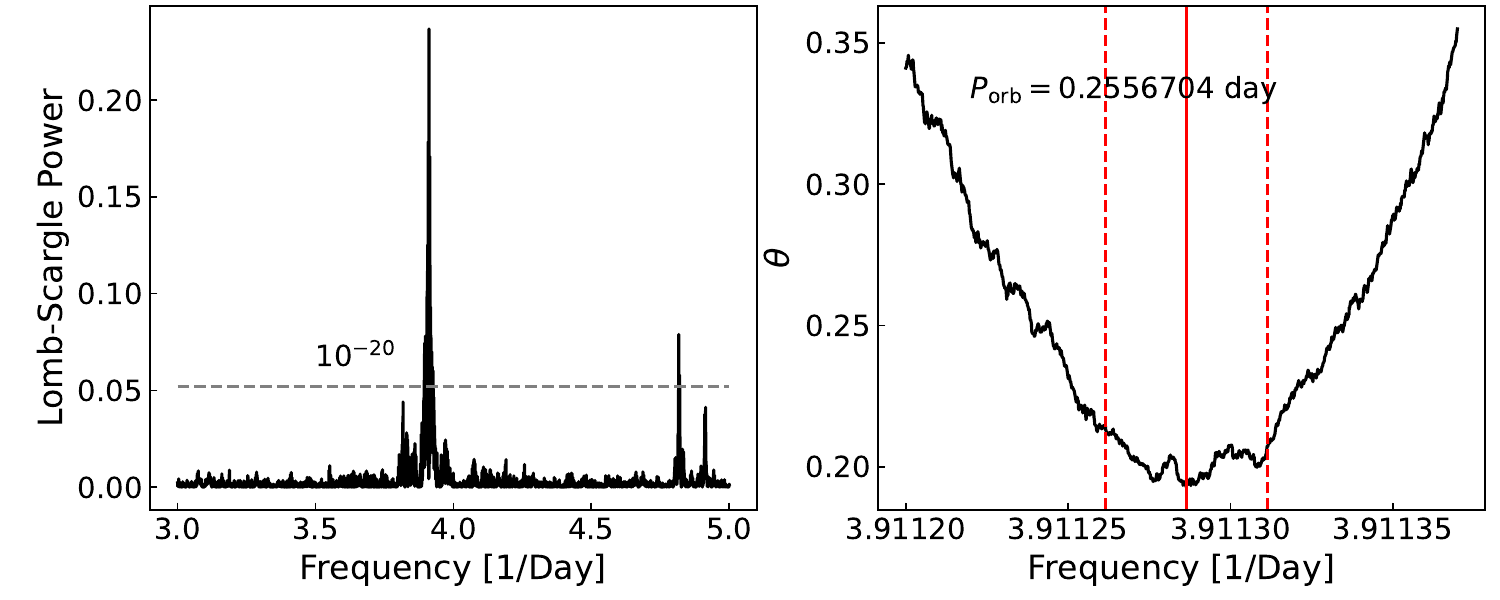}
\caption{Left panel: Power spectrum of TIC 16320250 obtained from the ASAS-SN, ZTF, and TESS light curves. A peak at frequency $\nu = 1/P \approx 3.9113535$ day$^{-1}$ ($P \approx$ 0.255666 day) can be clearly seen. Right panel: The phase dispersion calculated with these light curves, for the frequencies from 3.9112 day$^{-1}$ to 3.91137 day$^{-1}$, with a step of 10$^{-7}$ day$^{-1}$. The vertical red solid line shows the phase dispersion minimum.}
\label{period.fig}
\end{figure*}

\section{Long-term light curve and stellar cycle}
\label{lc_sc.sec}

TIC 16320250 has been observed by superWASP, Catalina, ASAS-SN, ZTF, and TESS, from 2004 to 2023 (Figure \ref{Vobs.fig}). No clear long-term periodic variation can be seen from the light curves.

\subsection{Orbital period measurement}
We analyzed the period with two independent techniques to derive an accurate orbital period (Figure \ref{period.fig}). The first technique is the Lomb–Scargle method \citep{1989ApJ...338..277P}, which is devised for unevenly spaced data. 
The light curves (LCs) from ASAS-SN, ZTF, and TESS were used to compute the Lomb periodogram. We found a significant peak abound $f =$ 3.91135 day$^{-1}$, with a probability $<$10$^{-20}$ that it is due to random fluctuations of photon counts. 
The second technique is the phase dispersion minimization (PDM) analysis \citep{1978ApJ...224..953S}. A search of periods in the frequency range of 3.91120--3.91137 day$^{-1}$ returns a phase dispersion minimum at $f =$ 3.9112864 day$^{-1}$, corresponding to a period of 0.2556704 day.

The folded LCs show the characteristic double-peaked morphology expected for a tidally distorted secondary (Figure \ref{lcs.fig}), with the two bright but asymmetrical peaks at $\phi =$ 0.25 and 0.75 suggesting a strong O'Connell effect (i.e., significant flux differences at quadrature phases) \citep{1951PRCO....2...85O}.
In addition, the gravity darkening normally leads to a fainter luminosity at the superior conjunction ($\phi =$ 0.5) than that at the inferior conjunction ($\phi =$ 0 or 1). 
However, for TIC 16320250, the phase with the faintest luminosity varies with different observations. 
All these indicate the existence of stellar spots.

\subsection{Stellar cycle}

\begin{figure*}                                                 
\center 
\includegraphics[width=1\textwidth]{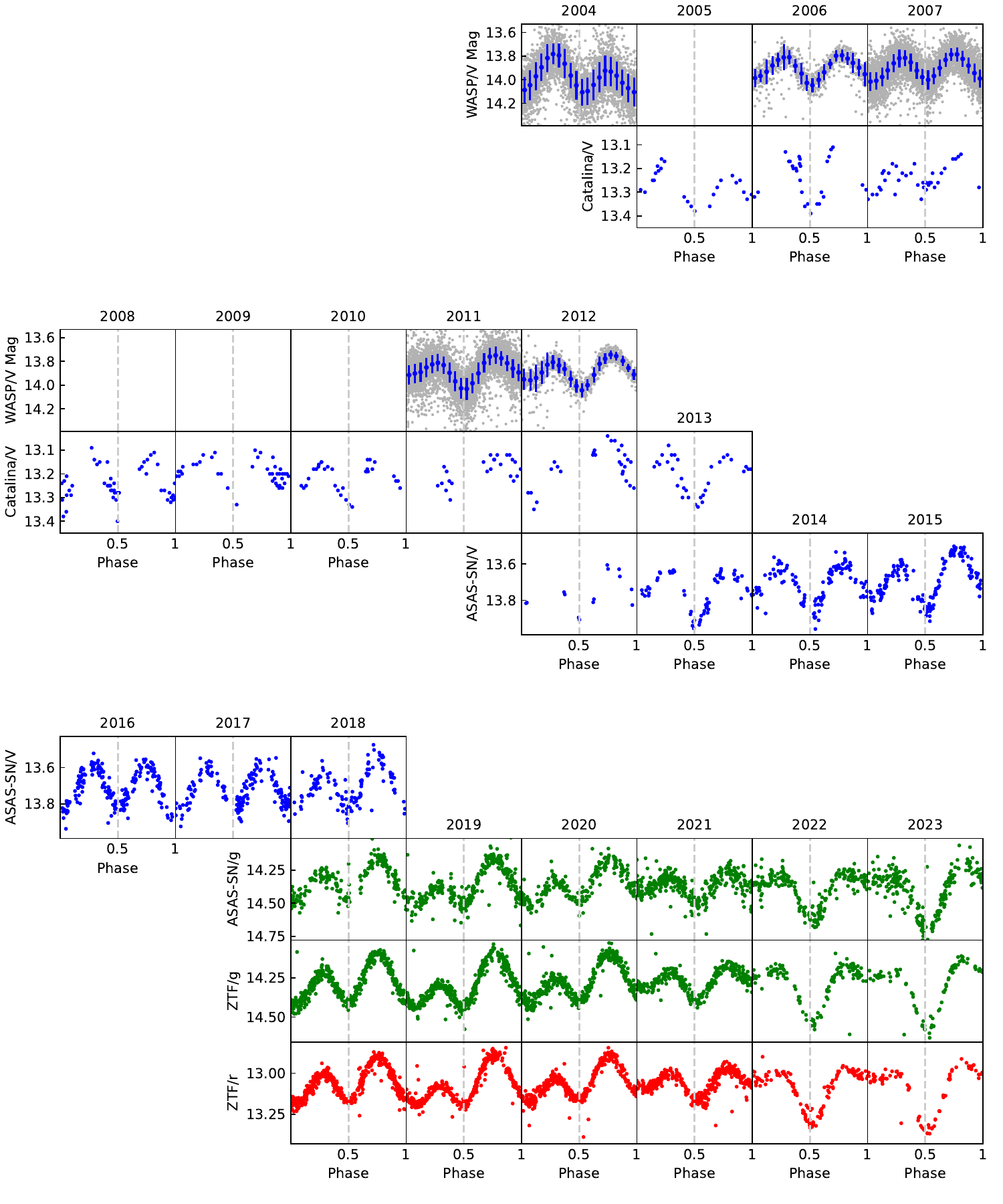}
\caption{Folded light curves in different bands ($V$ band of superWASP, Catalina, and ASAS-SN; $g$ band of ASAS-SN and ZTF; $r$ band of ZTF), with a period of 0.2556704 day. The phases are corrected to the inferior conjunction at $\phi =$ 0 or 1, quadrature at $\phi =$ 0.25 and 0.75, and superior conjunction at $\phi =$ 0.5 following radial velocity curve. The long-term light curves display three similar patterns, with deep dips occurring at a phase of 0.5, observed in 2006, 2014, and 2022.}
\label{lcs.fig}
\end{figure*}

\begin{figure*}[htbp!]
\center 
\includegraphics[width=1\textwidth]{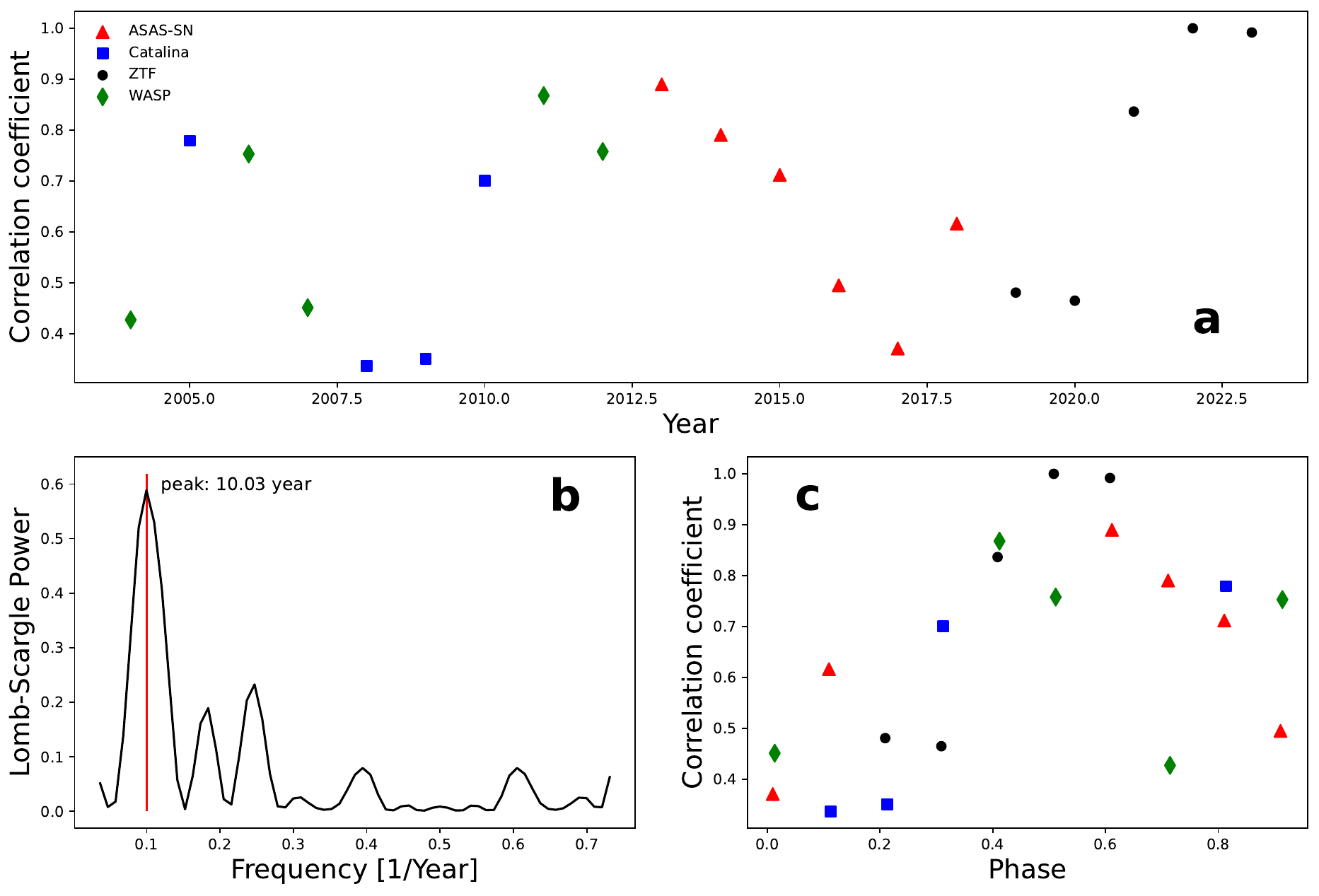}
\caption{Panel a: the correlation coefficients of $V$-band light curves in each year compared to ZTF $V$ band data ($=g-0.42\times(g-r)-0.03$) in 2022. Panel b: The Lomb-Scargle power spectrum of the correlation coefficients. The peak indicates a period of about 10.03 years. Panel c: the folded correlation coefficients using a period of 10.03 years.}
\label{ccf.fig}
\end{figure*}

The long-term light curves of TIC 16320250 spanning twenty years (2004--2023) display three analogous patterns, with deep dips occurring at a phase of 0.5, observed in 2006, 2014, and 2022 (Figure \ref{lcs.fig}).
This suggests a periodicity in the variation pattern of approximately 8 years.
Using the light curve from 2022 as the reference, we performed a cross-correlation analysis of these light curves in the $V$ band.
For the ZTF data, we converted the $g$- and $r$-band magnitudes into $V$ band with $V=g-0.42\times(g-r)-0.03$ \citep{1991ApJ...380..362W}.
Then the Lomb-Scargle method was applied to measure the period of the repeated pattern of the light curves.
This yields a period of about 10 years indicative of a potential activity cycle (Figure \ref{ccf.fig}).

\begin{figure}[htbp!]    
\center
\includegraphics[width=0.5\textwidth]{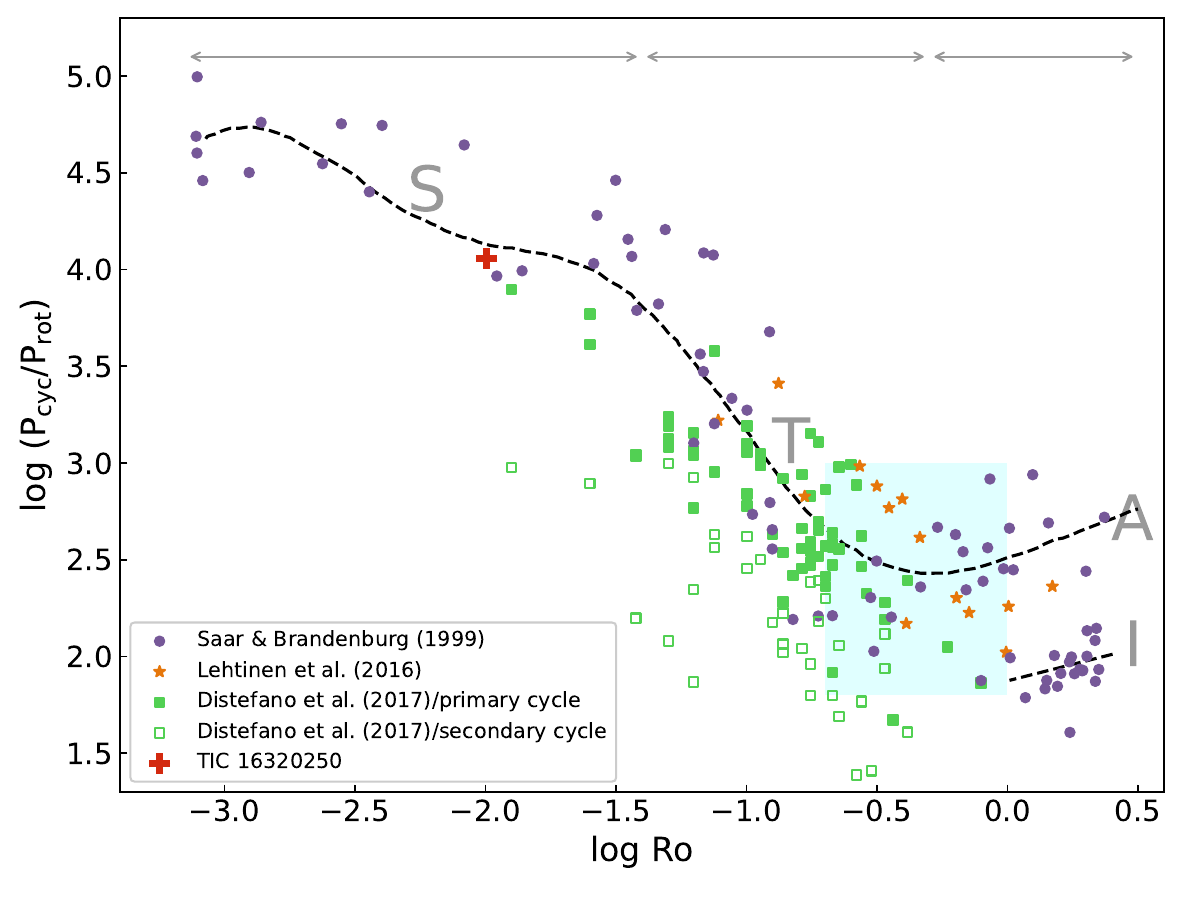}
\caption{Cycle frequency ($P_{\rm cyc}$/$P_{\rm rot}$) versus Rossby number (Ro = $P_{\rm rot}$/$\tau$). These points are located in four branches: superactive (S), transitional (T), active (A) and inactive (I). TIC 16320250 (red plus) lies on the superactive branch.}
\label{cycRo.fig}
\end{figure}

In order to confirm that the 10-year variation is due to stellar cycle, we plotted the diagram of cycle frequency ($P_{\rm cyc}$/$P_{\rm rot}$) vs. Rossby number (Ro $=P{\rm rot}$/$\tau_{\rm C}$), which divide stars into four branches: superactive, transitional, active and inactive branches, each of which may be related to a dynamo evolution sequence \citep{1999ApJ...524..295S}.
Note here Ro is equivalent to 4$\pi$ times Ro$_{\rm Br}$ used in previous studies \citep{1998A&A...340..437B}.
For TIC 16320250, the convective turnover time $\tau_{\rm C}$ was calculated by using the semi-empirical formula \citep{1984ApJ...287..769N} which gives $\tau_{\rm C}$ as a function of the 
$B-V$ color. 
TIC 16320250 lies on the superactive branch, further indicating our estimation of the cycle length is plausible (Figure \ref{cycRo.fig}), evidencing that the cycle searching method in this study is feasible.

The cycle observed in TIC 16320250 is distinct from the 11-year solar cycle, as the total brightness doesn't show a clear periodic variation in Figure \ref{Vobs.fig}.
However, it is well-established from previous studies that multiple stellar cycles exist in the sun and other distant stars.
Those multiple cycles inferred from light curves may indicate the variations of different spot properties \citep{2003AN....324..202R}.
Some systems show a Flip-flop cycle indicating two active longitudes about 180$^{o}$ apart on stellar surface, with alternating levels of spot activity in a cyclic manner \citep{1993A&A...278..449J}.
For close binaries, this could be interpreted as being due to a magnetic field connection between the two stars.
However, such a cycle is not observed in TIC 16320250.

\section{Orbital solution}
\label{orbit.sec}

\subsection{Radial Velocity fitting}

We measured RV with the classical cross-correlation technique, 
i.e. by shifting and comparing the best-matched template to the observed spectrum. The PHOENIX model with the parameter of $T_{\mathrm{eff}}=4300$ K, $\mathrm{log}\textit{g}=4.5$ and $[\mathrm{Fe/H}]=-0.5$ was used as the template.
The RV grid added to the template is $-$400 km/s $\sim$ 400 km/s with a step of 0.5 km/s.
The $\chi^2$ curve was fitted with a Gaussian function to find the minimum value, corresponding to the final RV, and the fitting uncertainty.
The error of the final RV is the square root of the sum of the squares of 
the $\sigma$ of the Gaussian function and the formal error where $\Delta\chi^2=1$. Table \ref{rvdata.tab} lists the RV data from LAMOST and 2.16 m observations.

\begin{figure}[h!]   
\center 
\includegraphics[width=0.48\textwidth]{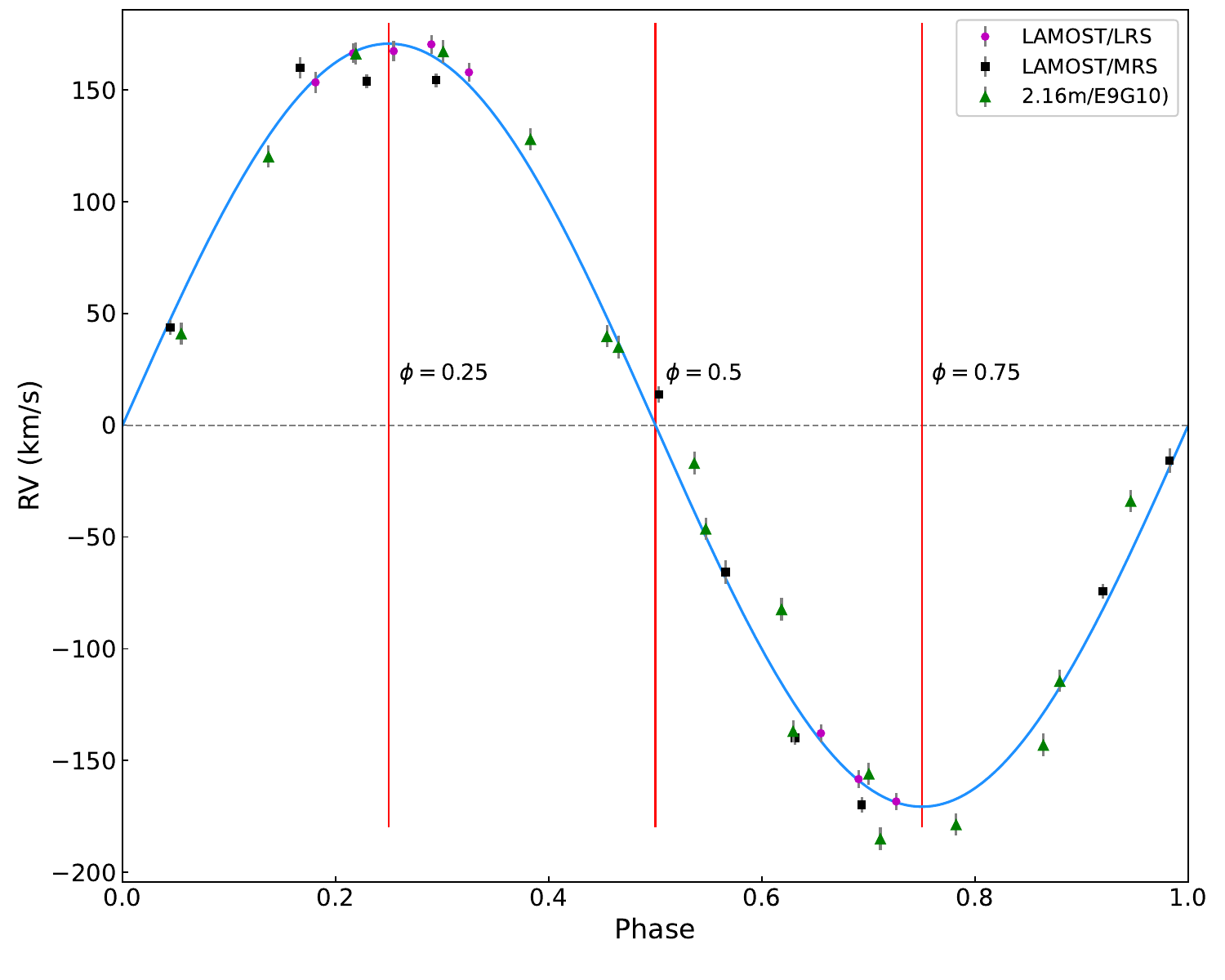}
\caption{Folded radial velocity curve with a period of 0.2556704 day. The phases of the inferior conjunction ($\phi =$ = 0 or 1), quadrature ($\phi =$ 0.25 and 0.75), and superior conjunction  $\phi =$ 0.5 are marked with vertical red lines. The RV data are determined from the LAMOST low-resolution spectra (violet circles), LAMOST med-resolution $B$-band spectra (black squares), and 2.16m E9+G10 spectra (green triangles).}
\label{fold.fig}
\end{figure}

{\it The Joker} \citep{2017ApJ...837...20P}, a python module, can fit the radial velocity well by using a Markov chain Monte Carlo sampler. The fitted orbital parameters from {\it The Joker} are: 
period $P= 0.255668^{+0.000001}_{-0.000001}$ day,
eccentricity $e = 0.09^{+0.02}_{-0.02}$, 
argument of the periastron $\omega = 1.99^{+0.14}_{-0.12}$, 
mean anomaly at the first exposure $M{\rm 0} = -0.34^{+0.13}_{-0.12}$, 
semi-amplitude $K = 173^{+2}_{-2}$ km/s, 
and systematic RV $\nu{\rm 0} = -23^{+2}_{-2}$ km/s. Figure \ref{fold.fig} shows the folded radial velocity data and the fit curve (with the period of 0.2556704 day).

Using those fit parameters, we calculated the mass function
following
\begin{equation}
    f(M) = \frac{M_{2} \, \textrm{sin}^3 i} {(1+q)^{2}} = \frac{P \, K_{1}^{3} \, (1-e^2)^{3/2}}{2\pi G},
\label{formula2}
\end{equation}
\noindent
where $M_{2}$ is the mass of the secondary in binary system, $q = M_{1}/M_{2}$ is the mass ratio, and $i$ is the system inclination.
The mass function is $f(M) = 0.135^{+0.005}_{-0.005}\ M_{\odot}$.

\subsection{LC fitting}
\label{lcfit.sec}

\subsubsection{Unspotted brightness}

The unspotted brightness is important for the light curve fitting with stellar spots (especially polar spots) since the unspotted brightness and the spot parameters are degenerate.
In previous studies, the unspotted brightness was estimated from light curve fitting or some novel methods.
For example, \citet{1992MNRAS.259..302O,1997A&A...321..811O} found a nearly linear relation between the brightness of the star and the light curve amplitude or Mg II h\&k line fluxes: larger light curve amplitude corresponds to lower brightness at minimum light, while higher Mg II flux corresponds to lower stellar brightness.
Therefore, the unspotted brightness can be estimated when the amplitude of the light curve or Mg II h\&k line fluxes decreases to zero.
This method needs long-term precise photometric or spectroscopic observations to cover the full variation range of light curves or Mg II h\&k line fluxes.
However, for TIC 16320250, the ZTF data spans only seven years and the amplitudes of light curves are mostly large values, which results in a poor linear fitting of the relation.

On the other hand, we found that a long-term joint fitting (with the unspotted brightness as a free but shared parameter for each year) can break the degeneracy and return a true unspotted brightness.
As a test, we built a similar semi-detached binary system and set the intrinsic luminosity of the visible star to \emph{pblum}=15 in PHOEBE \citep{2016ApJS..227...29P, 2018ApJS..237...26H, 2020ApJS..250...34C}, and produced five light curves with different spot parameters, as shown in Table \ref{theo_lcs.tab}. 
We performed fittings to these light curves in two ways: a simultaneous fitting with shared unspotted brightness and a separate fitting with individual unspotted brightness.
Table \ref{test_results.tab} shows the fitting results. 
The simultaneous fitting returns an unspotted value close to the true value (\emph{pblum}=15), while the separate fittings return different and inaccurate unspotted values for each light curve.
Therefore, we can derive approximate estimations of unspotted brightness using simultaneous fitting to long-term light curves covering different spot parameters.

\begin{table}
\caption{Five faked light curves with different spot parameters. The $radius$, $relteff$, $colat$ and $long$ represent the angular radius, the ratio of temperature, colatitude, and longitude of the spot, respectively.
\label{theo_lcs.tab}}
\setlength{\tabcolsep}{4mm}
\centering
 \begin{tabular}{cccccc}
\hline
Parameter & 1 & 2 & 3 & 4 & 5  \\
\hline
$radius \ $ ($^{\circ}$) & 40 & 45 & 35 & 40 & 35  \\
$relteff$ & 0.8 & 0.8 & 0.8 & 0.8 & 0.8  \\
$colat \ $ ($^{\circ}$) & 5 & 10 & 15 & 10 & 10  \\
$long \ $ ($^{\circ}$) & 120 & 90 & 60 & 30 & 60  \\
\hline
\end{tabular}
\end{table}

\begin{table}
\caption{The value of the intrinsic luminosity by fitting faked light curves simultaneously and separately. $pblum_{\rm sim}$ and $pblum_{\rm sep}$ represent the intrinsic luminosity obtained by simultaneous and separate fitting, respectively.
\label{test_results.tab}}
\setlength{\tabcolsep}{0.5pt}
\centering
 \begin{tabular}{cccccc}
\hline
Parameter & 1 & 2 & 3 & 4 & 5  \\
\hline
$pblum_{\rm sim}$ &  &  & $14.78_{-0.34}^{+0.29}$ &  &   \\
$pblum_{\rm sep}$ & $14.44_{-0.54}^{+0.77}$ & $14.09_{-0.63}^{+0.93}$ & $14.88_{-0.43}^{+0.62}$ & $14.36_{-0.52}^{+0.85}$ & $14.84_{-0.48}^{+0.66}$  \\
\hline
\end{tabular}
\end{table}

\subsubsection{Mass and radius as initial input parameters}
\label{pmass.sec}

First, we calculated the mass and radius of the M star assuming a single-star evolution.
The stellar spectroscopic mass also can be estimated by the following formula:
\begin{equation}
M = \frac{L}{4\pi~G\sigma~T^{4}}g.  
\label{formula1}
\end{equation}
We used multi-band magnitudes ($G$, $G_{\rm BP}$, $G_{\rm RP}$, $J$, $H$, and $K_{\rm S}$) to calculate the bolometric magnitude. No absorption correction was applied due to the small extinction value ($\approx$0). 
With the absolute luminosity and magnitude of the sun ($L_{\odot} =$ 3.83$\times$ 10$^{33}$ erg/s; $M_{\odot} =$ 4.74), the bolometric luminosity was calculated following,
\begin{equation}
M_{\odot} - M_{\rm bol} = 2.5\log\frac{L_{\rm bol}}{L_{\odot}}. 
\end{equation} 
We derived the spectroscopic stellar mass and radius of $M\ =\ 0.62\pm0.01$ M$_{\odot}$ and $R\ =\ 0.63\pm0.01$ R$_{\odot}$.
In addition, the {\it isochrones} module returned evolutionary mass and radius estimations of $M\ =\ 0.60^{+0.02}_{-0.02}$ M$_{\odot}$ and $R\ =\ 0.68^{+0.01}_{-0.01}$ R$_{\odot}$, consistent with the spectroscopic estimations.

Second, the $H_{\alpha}$ emission line profiles show a component from the disk around the unseen object (Section \ref{activity.sec}), implying the M star has filled its Roche lobe and mass transfer has started.
For a low-mass main-sequence star which (almost) fills its Roche lobe, the lobe radius can be calculated with the period-radius relation following
\begin{equation}
    \label{eq:rhostar}
R \approx 0.11P_{\rm orb}{\rm (hr)} R_{\odot} \approx 0.67 R_{\odot}.
\end{equation}
Its mean density can be calculated following \citep{2002apa..book.....F},
\begin{equation}
\bar{\rho} = 110P^{-2}_{\rm hr}\ {\rm g}\ {\rm cm^{-3}}.
\end{equation}
Using this mean density and the lobe radius, the M star has a mass of $0.62^{+0.03}_{-0.03} M_{\odot}$, similar to the mass estimated from single-star evolution.
This means the mass transfer may have just started for a short time.

\subsubsection{LC fitting to long-term light curves}

This system has been observed for many years by superWASP, Catalina, ASAS-SN, TESS and ZTF. Figure \ref{lcs.fig} shows its light curves from 2004 to 2023 in different bands. Besides ellipsoidal modulation, the light curves show clear variation caused by cool spots or warm faculae. Here we used PHOEBE \citep{2016ApJS..227...29P, 2018ApJS..237...26H, 2020ApJS..250...34C} to simultaneously fit the multi-year light curves of this system by adding one or two cool spots to the visible star.
We used the light curves of ZTF $r$ band and $g$ band from 2018 to 2023, TESS data in 2020 and $B$/$V$/$R$ data from 85 cm observations in 2022. 
In the light curve modeling, we set the unseen star as a small ($3 \times 10^{-6} R_{\odot}$) and cold (300 K) blackbody, and used the {\it eclipse\_method = only\_horizon} as the eclipse model. 
The binary system was set as a semi-detached system since the visible star (almost) fills its Roche lobe. 
The orbital parameters from RV fitting ($P$ =0.25567 day, $e$ = 0 and $K$ = 172.6 km/s) were fixed.

\begin{figure*}[htbp!]
\center 
\includegraphics[width=1\textwidth]{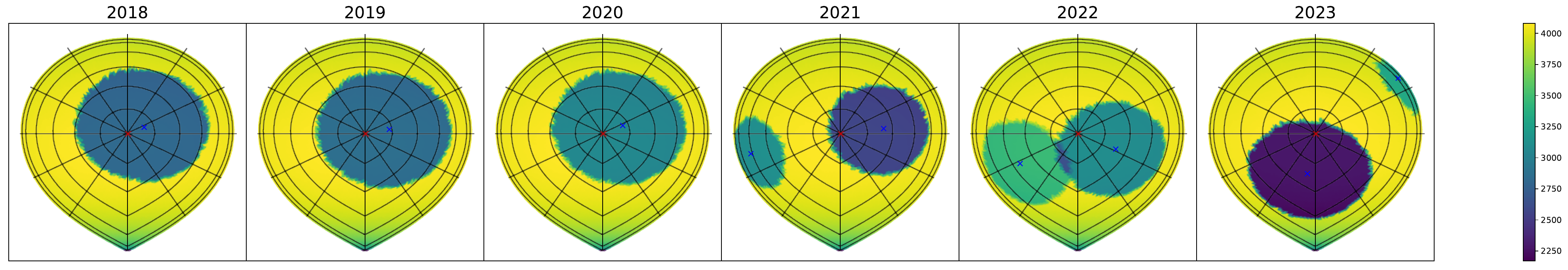}
\includegraphics[width=1\textwidth]{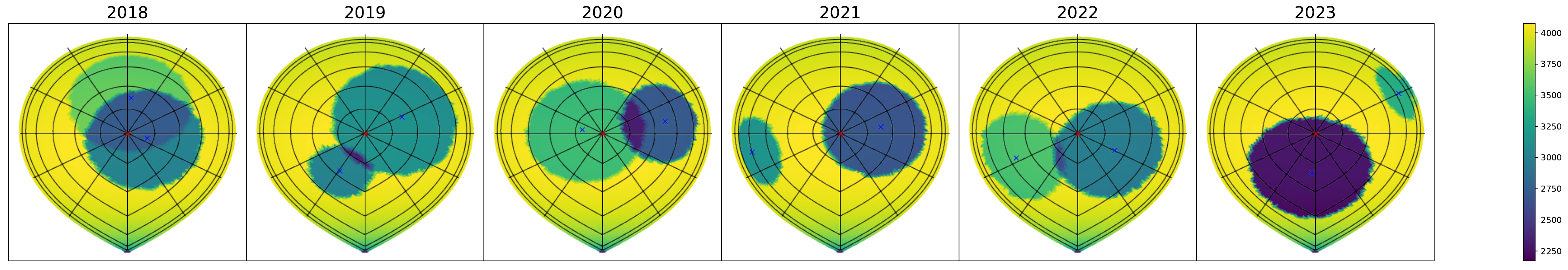}
\caption{Size and configurations of spots from PHOEBE fitting. The red cross marks the center of the pole, and the blue crosses show the centers of spots. Top panel: PHOEBE fitting of model A. Bottom panel: PHOEBE fitting of model B.}
\label{spot_fig.fig}
\end{figure*}

During the fitting, we found that the light curves from 2021 to 2023 can not be fit well with one single spot.
Therefore, we tried two models: model A, using a single-spot model for the light curves from 2018 to 2020 and a two-spots model for light curves from 2021 to 2023; model B using two spots for each year.
Figure \ref{spot_fig.fig} plots the sizes and configurations of the stellar spots for both two models, which shows clear variation of a large polar spot or a gathering of small spots in the polar region.
Figure \ref{fit_lcs.fig} and \ref{fit_lcs_2spots.fig} show the best-fit model using these two models.
For the TESS light curve in 2020, the fit result of model B is better than that of model A since the former fitting has a smaller reduced $\chi^2$.
Table \ref{phoebe.tab} lists the PHOEBE fitting results for the system, the primary (i.e., the visible star) and the secondary (i.e., the unseen star), while Table \ref{spots.tab} lists the parameters of spots.
The two models return similar results of the orbital solution and stellar properties.
Take model B as an example, the PHOEBE fitting returns a binary system with an orbital inclination angle being about $61.80^{\circ}$$^{+0.36^{\circ}}_{-0.52^{\circ}}$ and a mass ratio ($secondary/primary$) being about $1.20^{+0.01}_{-0.01}$. 
The mass of visible star is about $0.56^{+0.02}_{-0.02}\ M_{\odot}$, and the unspotted brightness in ZTF $r$ and $g$ band are $12.84^{+0.01}_{-0.01}$ mag and $14.04^{+0.01}_{-0.01}$ mag, respectively.
The mass of the unseen object is $0.67^{+0.02}_{-0.02} M_{\odot}$, which is significantly smaller than the Chandrasekhar limit. 
Therefore, it is most likely a white dwarf.

\begin{table}
\caption{The PHOEBE parameter estimates for this system, primary and secondary star by using different models. \label{phoebe.tab}}
\centering
\setlength{\tabcolsep}{1mm}
 \begin{tabular}{cccc}
\hline
Parameter & system & primary & secondary \\
\hline
\multicolumn{4}{c}{model A}\\
\hline
$P_{\rm orb}$ (${\rm d}$) & 0.25567 (fixed) \\
$e$ & 0 (fixed) \\
$i$ ($ \ ^{\circ}$) & $61.02^{+0.51}_{-0.91}$ \\
$q$ & $1.22^{+0.01}_{-0.01}$ \\
r band (mag) & & $12.83^{+0.01}_{-0.01}$ \\
g band (mag) & & $14.04^{+0.01}_{-0.01}$ \\
$a{\rm sin}i$ (${\rm R}_{\odot}$) & & 0.872 (fixed) & $0.71^{+0.01}_{-0.01}$ \\
$R$ (${\rm R}_{\odot}$) & & $0.66^{+0.01}_{-0.01}$ & $3 \times 10^{-6}$ (fixed) \\
$T_{\rm eff}$ (${\rm K}$) & & $3971.0^{+5.2}_{-9.4}$ & 300 (fixed) \\
$M$ (${\rm M}_{\odot}$) & & $0.55^{+0.03}_{-0.03}$ & $0.67^{+0.03}_{-0.03}$ \\
\hline
\multicolumn{4}{c}{model B}\\
\hline
$P_{\rm orb}$ (${\rm d}$) & 0.25567 (fixed) \\
$e$ & 0 (fixed) \\
$i$ ($ \ ^{\circ}$) & $61.80^{+0.36}_{-0.52}$ \\
$q$ & $1.20^{+0.01}_{-0.01}$ \\
r band (mag) & & $12.84^{+0.01}_{-0.01}$ \\
g band (mag) & & $14.04^{+0.01}_{-0.01}$ \\
$a{\rm sin}i$ (${\rm R}_{\odot}$) & & 0.872 (fixed) & $0.73^{+0.01}_{-0.01}$ \\
$R$ (${\rm R}_{\odot}$) & & $0.66^{+0.01}_{-0.01}$ & $3 \times 10^{-6}$ (fixed) \\
$T_{\rm eff}$ (${\rm K}$) & & $3973.4^{+9.1}_{-2.4}$ & 300 (fixed) \\
$M$ (${\rm M}_{\odot}$) & & $0.56^{+0.02}_{-0.02}$ & $0.67^{+0.02}_{-0.02}$ \\
\hline
\end{tabular}
\end{table}

\begin{table*}
\caption{PHOEBE parameter estimates for spots of model A and B, including $radius$, $relteff$, $colat$ and $long$ representing the angular radius, the ratio of temperature, colatitude, and longitude of spot, respectively. The subscripts 1 and 2 represent the first and second spots, respectively. \label{spots.tab}}
\centering
 \begin{tabular}{ccccccc}
\hline
Parameter & 2018 & 2019 & 2020 & 2021 & 2022 & 2023 \\
\hline
\multicolumn{7}{c}{model A}\\
\hline
$radius_{\rm 1} \ $ ($^{\circ}$) & $40.54^{+0.21}_{-0.18}$ & $41.69^{+0.27}_{-0.41}$ & $40.78^{+0.19}_{-0.29}$ & $31.65^{+1.69}_{-1.00}$ & $34.04^{+1.33}_{-0.82}$ & $20.22^{+0.88}_{-0.76}$ \\
$relteff_{\rm 1}$ & $0.69^{+0.02}_{-0.01}$ & $0.70^{+0.01}_{-0.01}$ & $0.75^{+0.01}_{-0.01}$ & $0.63^{+0.05}_{-0.07}$ & $0.76^{+0.01}_{-0.01}$ & $0.83^{+0.05}_{-0.04}$ \\
$colat_{\rm 1} \ $ ($^{\circ}$) & $12.93^{+0.31}_{-0.17}$ & $14.14^{+0.14}_{-0.14}$ & $12.97^{+0.19}_{-0.24}$ & $25.76^{+1.58}_{-2.80}$ & $25.16^{+1.25}_{-3.50}$ & $79.84^{+4.45}_{-6.63}$ \\
$long_{\rm 1} \ $ ($^{\circ}$) & $125.70^{+0.44}_{-0.24}$ & $105.88^{+0.74}_{-0.80}$ & $116.31^{+1.21}_{-0.51}$ & $99.15^{+0.92}_{-0.89}$ & $64.19^{+7.14}_{-7.33}$ & $125.44^{+2.14}_{-3.61}$ \\
\hline
$radius_{\rm 2} \ $ ($^{\circ}$) & & & & $24.23^{+1.02}_{-0.94}$ & $30.04^{+1.16}_{-2.47}$ & $36.42^{+0.47}_{-0.23}$ \\
$relteff_{\rm 2}$ & & & & $0.77^{+0.01}_{-0.01}$ & $0.85^{+0.03}_{-0.02}$ & $0.56^{+0.03}_{-0.05}$ \\
$colat_{\rm 2} \ $ ($^{\circ}$) & & & & $60.24^{+2.29}_{-4.86}$ & $41.70^{+1.89}_{-2.78}$ & $27.09^{+0.26}_{-2.77}$ \\
$long_{\rm 2} \ $ ($^{\circ}$) & & & & $285.32^{+0.70}_{-2.39}$ & $301.12^{+3.23}_{-3.17}$ & $351.07^{+1.15}_{-0.67}$ \\
\hline
\multicolumn{7}{c}{model B}\\
\hline
$radius_{\rm 1} \ $ ($^{\circ}$) & $35.15^{+0.49}_{-0.13}$ & $24.12^{+6.39}_{-6.55}$ & $27.10^{+0.21}_{-0.06}$ & $32.99^{+0.18}_{-0.14}$ & $34.04^{+0.79}_{-0.78}$ & $19.55^{+1.16}_{-1.45}$ \\
$relteff_{\rm 1}$ & $0.72^{+0.03}_{-0.02}$ & $0.82^{+0.06}_{-0.08}$ & $0.67^{+0.01}_{-0.01}$ & $0.66^{+0.01}_{-0.01}$ & $0.73^{+0.04}_{-0.02}$ & $0.84^{+0.03}_{-0.04}$ \\
$colat_{\rm 1} \ $ ($^{\circ}$) & $11.54^{+0.40}_{-0.70}$ & $20.77^{+10.58}_{-4.36}$ & $39.01^{+0.08}_{-0.31}$ & $23.57^{+0.46}_{-0.34}$ & $24.00^{+2.85}_{-0.81}$ & $71.91^{+10.15}_{-3.46}$ \\
$long_{\rm 1} \ $ ($^{\circ}$) & $73.87^{+5.85}_{-2.43}$ & $-23.21^{+12.34}_{-12.46}$ & $103.84^{+0.51}_{-0.73}$ & $101.49^{+0.60}_{-0.80}$ & $61.86^{+3.35}_{-1.87}$ & $121.88^{+2.58}_{-4.64}$ \\
\hline
$radius_{\rm 2} \ $ ($^{\circ}$) & $38.44^{+1.97}_{-2.64}$ & $39.80^{+1.18}_{-0.66}$ & $35.85^{+0.08}_{-0.14}$ & $23.88^{+1.65}_{-0.74}$ & $30.56^{+3.71}_{-0.99}$ & $36.98^{+0.48}_{-0.35}$ \\
$relteff_{\rm 2}$ & $0.92^{+0.01}_{-0.02}$ & $0.78^{+0.02}_{-0.01}$ & $0.85^{+0.01}_{-0.01}$ & $0.78^{+0.05}_{-0.04}$ & $0.87^{+0.01}_{-0.01}$ & $0.56^{+0.04}_{-0.02}$ \\
$colat_{\rm 2} \ $ ($^{\circ}$) & $25.71^{+1.62}_{-0.90}$ & $23.01^{+1.40}_{-0.77}$ & $11.87^{+0.23}_{-0.04}$ & $61.28^{+1.12}_{-1.32}$ & $41.73^{+2.80}_{-5.65}$ & $24.74^{+1.05}_{-2.33}$ \\
$long_{\rm 2} \ $ ($^{\circ}$) & $177.07^{+2.12}_{-1.82}$ & $119.37^{+0.90}_{-2.52}$ & $256.81^{+0.85}_{-1.45}$ & $283.62^{+0.87}_{-0.74}$ & $294.35^{+2.91}_{-2.93}$ & $352.77^{+1.74}_{-1.31}$ \\
\hline
\end{tabular}
\end{table*}

\section{Discussion}
\label{discuss.sec}

\subsection{Stellar cycle and Evolution of Polar spots}

The variation of the light curves during one cycle ($\approx 10$ years) is most likely due to the evolution of spots.
By performing a joint fitting to the long-term multi-band light curves, we determined the unspotted brightness and the parameters of the spots.
The fitting with two spots yields better results than the fitting with a single spot, and reveals a possible motion of the spots during these years (Figure \ref{spot_fig.fig}).

The high-latitude or polar starspots in TIC 16320250 have been observed in many rapidly-rotating stars, either from Doppler imaging \citep{1992ASPC...26..249K,1999ApJS..121..547V, 1999A&A...347..225S} or light curve modeling \citep{1997A&A...321..811O, 2001A&A...372..119O}.
These spots may cover a large fraction of the stellar photosphere and typically have lower temperatures, larger areas and longer lifetimes than those on low latitudes \citep{1996A&A...316..164R}.
In RS CVn binaries or young main-sequence stars, the lifetime of polar spots can be over a decade \citep{2009A&ARv..17..251S,2019dmde.book.....H}.  
These starspots may be a result of strong Coriolis force acting on magnetic flux tubes that rise from deep regions within the star \citep{1992A&A...264L..13S}, or they may initially appear at low latitudes and then be advected polewards by near-surface meridional flows \citep{2001ApJ...551.1099S}.
Enhanced magnetic flux by unstable magnetic Rossby waves at high latitudes of tachoclines may also lead to the formation of polar spots \citep{2011A&A...532A.139Z}.
Additionally, a mechanism based on a self-consistent distributed dynamo has been proposed for the formation of sizable high-latitude dark spots \citep{2015A&A...573A..68Y}.

Although the mechanisms for generating polar spots have been widely studied, their motion on stellar surface and their lifetimes and cycles, are rarely investigated. 
Only few stars have had their polar spots' motion studied by photometric observations \citep{2001A&A...372..119O,2003AN....324..202R} or Doppler images \citep{1999ApJS..121..547V}. 
On the one hand, this is due to the difficulty of constraining the inclination angle of a single star, which is essential in determining the spot's location, through photometric observations.
On the other hand, although Doppler imaging can give a rough estimate of inclination angle \citep{1989ApJ...341..456H},
long-term high-resolution spectroscopic observations are time-consuming.
However, in our work, the joint fitting of long-term, multi-band light curves offers a way to measure the motion of spots, since the inclination angle of the visible can be measured assuming stellar rotation and binary orbit are coplanar.

The mechanism responsible for the motion of polar spots may be linked to the stellar magnetic cycle and magnetic dynamo.
The migration of high-latitude spots in II Peg was explained to be caused by a shorter rotation period than the orbital period, meaning either non-synchronous rotational and orbital periods or differential rotation \citep{1998A&A...340..437B}.
This is not the case of TIC 16320250, for which the rotation has been synchronized to the orbital motion.
Global dynamo models predict that the presence of strong magnetic fields in rapidly rotating low-mass stars leads to suppressed or quenched differential rotation by connecting different regions within the star's interior \citep{2013A&A...549L...5G}.
Thus, the polar spot motion in TIC 16320250 can not be driven by differential rotation.
The continuous changes in the positions and sizes of polar spots are similar to the RS CVn variables HK Lacertae \citep{1997A&A...321..811O} and IM Pegasi \citep{2003AN....324..202R}. 
In IM Pegasi, the variation timescales of the radius and longitude of one polar spot (29.8 and 10.4 years, respectively) agree well with the brightness cycle lengths of 28.2 and 10.1 years \citep{2003AN....324..202R}. 
In contrast, the continuous Doppler imaging of another RS CVn variable EI Eri over 11 years revealed no significant areal changes of its huge cap-like polar spot, neither did the observations for HR 1099 \citep{2009A&ARv..17..251S}.

\subsection{The nature of companion}
\label{nat.sec}

TIC16320250 has been reported as a neutron star with a mass of 0.98 $M_{\odot}$ \citep{2023ApJ...944L...4L}.
The authors derived orbital parameters and spot properties by individually fitting the light curves in each year.
That implies the unspotted brightness would be different in each fitting. 
Consequently, this could lead to inaccurate estimations of the orbital solution, particularly the inclination angle and mass ratio, and spot properties, given their degeneracy.

\begin{figure}[htbp!]   
\center 
\includegraphics[width=0.48\textwidth]{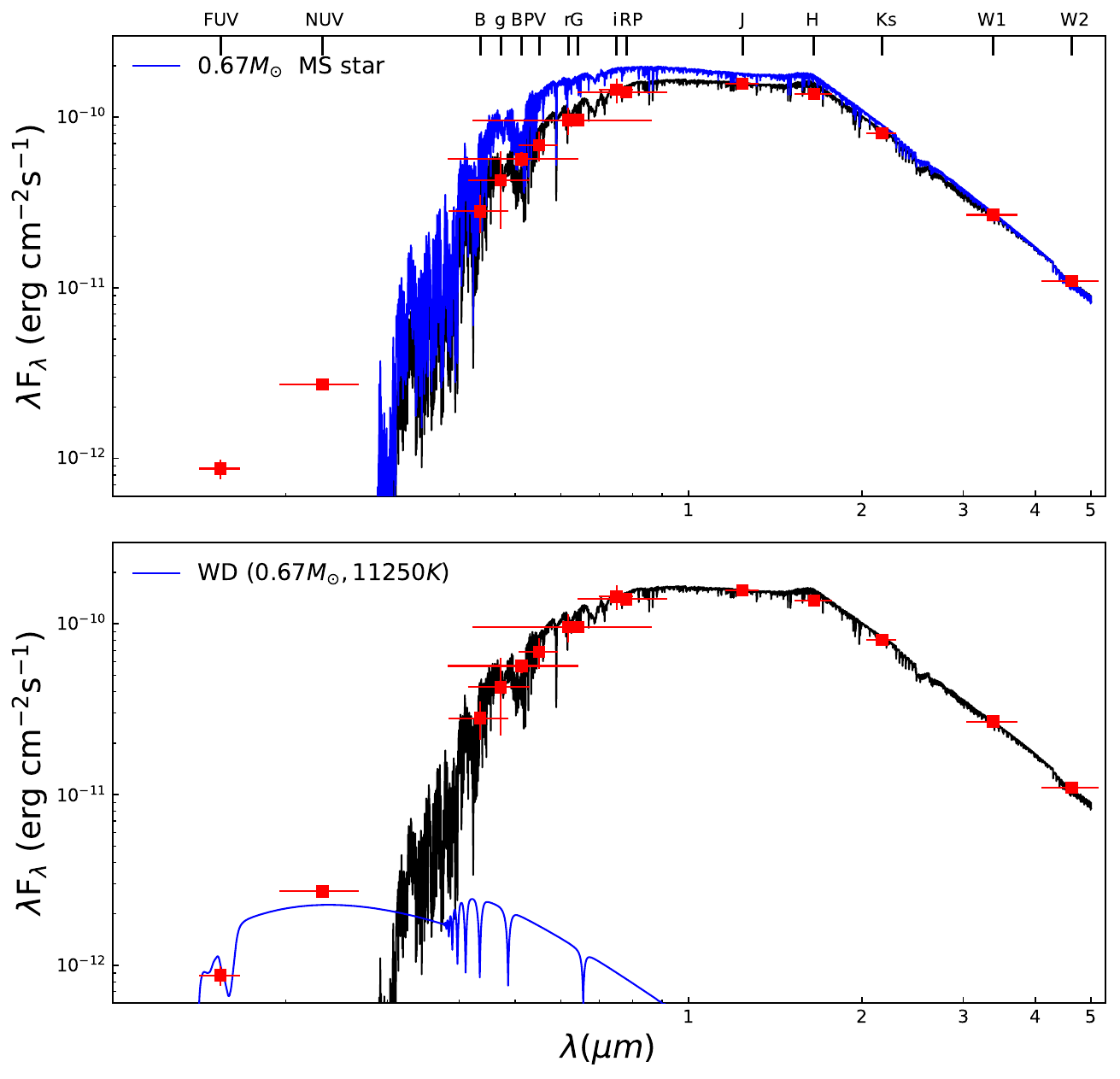}
\caption{Top panel: SED of TIC 16320250. The red squares are the observed photometry. The black line is the best-fit model to the photometry and the blue line is the spectrum of a main-sequence star with a mass of 0.67
$M_{\odot}$. Bottom panel: the blue line is the spectrum of a white dwarf with a mass of 0.67
$M_{\odot}$ and an effective temperature of 11250 $K$.}
\label{wdspectrum.fig}
\end{figure}

The $G$-band absolute magnitude of the visible star is 7.70 mag, while the absolute magnitude in $G$ band is 7.02 mag for a normal star with a mass of 0.67 $M_{\odot}$.
Thus the secondary can not be a normal main-sequence star since it would be brighter than the primary star (Figure \ref{wdspectrum.fig}.
The companion is most likely a white dwarf with a mass of 0.67 $M_{\odot}$.
The FUV and NUV emissions can be used to set an upper limit on the white dwarf's surface temperature, assuming the UV emissions are totally from the white dwarf.
A comparison with white dwarf models \citep{2010MmSAI..81..921K} indicates that a white dwarf with an effective temperature exceeding 11250 K can be conclusively excluded (Figure \ref{wdspectrum.fig}).

\subsection{Multiple stellar activity indicators}
\label{activity.sec}

TIC 16320250 (1RXS J152748.8+353658) has been observed by ROSAT telescope, with the count rate being 0.03$\pm$0.01 s$^{-1}$ and hardness ratio being 0.62$\pm$0.31.
The X-ray emission was thought as the result of stellar activity \citep{2011ApJ...743...48W, 2018MNRAS.476.1224A,2020A&A...638A..20M}.
By using the Gaia DR2 distance ($\approx$118 pc), the X-ray luminosity was determined as $L_X = $ 6.17$\times$10$^{29}$ erg/s and the X-ray activity was calculated as $R_X = \frac{L_X}{L_{\rm bol}}=-$2.79$\pm$0.10 \citep{2020A&A...638A..20M} .

We also calculated the classical chromospheric activity indicators $S$ index
 and $R^{'}_{HK}$ by using the Ca H$\&$K emission lines.
By using the multiple LAMOST low-resolution spectra, the $S$ index is determined to be 5.58 to 9.07, and the $R^{'}_{HK}$ is from $-$3.81 to $-$4.51.

Furthermore, we calculated the ultraviolet activity using the GALEX FUV and NUV bands. In brief, we calculated the UV activity index as follows:
\begin{equation}                                                   
R^\prime_{\rm UV} = \frac{f_{\rm UV,exc}}{f_{\rm bol}} = \frac{f_{\rm UV,obs} - f_{\rm UV,ph}}{f_{\rm bol}},
\label{eq:act_index}                                               
\end{equation}                                                     
where `UV' stands for the NUV and FUV bands, respectively.         
Here $f_{\rm UV,exc}$ is the UV excess flux due to activity.       
The observed UV flux $f_{\rm UV,obs}$ was estimated from the {\it GALEX} magnitude, while the photospheric flux $f_{\rm UV,ph}$, which means the photospheric contribution to the FUV and NUV emission,        
were estimated with the BTSettl model.
The bolometric flux $f_{\rm bol}$ was obtained with the effective temperature as $\sigma T^{4}_{\rm eff}$.
The $R^{'}_{\rm FUV}$ and $R^{'}_{\rm NUV}$ are calculated as $-$2.89 and
$-$2.2, respectively.

\begin{figure*}[htbp!]    
\center 
\includegraphics[width=1\textwidth]{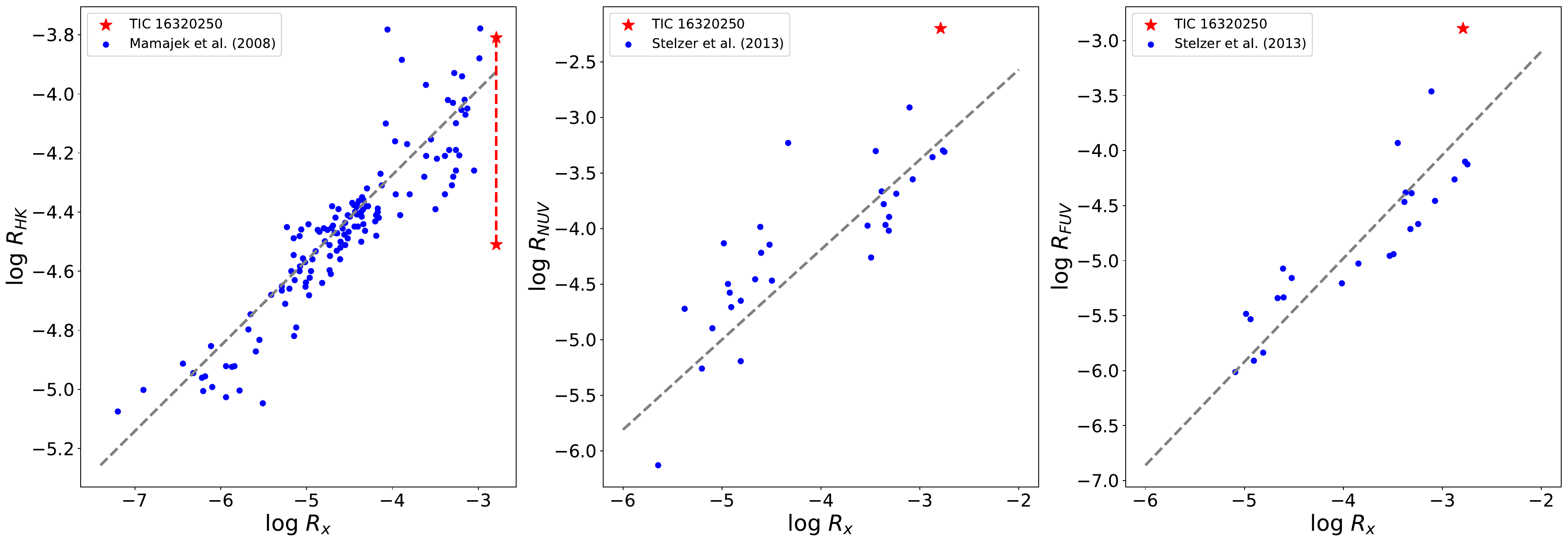}
\caption{Left panel: comparison between log$R_{X}$ and log$R_{HK}$. The blue points are from \citet{2008ApJ...687.1264M}. Middle panel: comparison between log$R_{X}$ and log$R_{NUV}$. The blue points are from \citet{2013MNRAS.431.2063S}. Right panel: comparison between log$R_{X}$ and log$R_{FUV}$.}
\label{activity.fig}
\end{figure*}

Previous studies reported a wrong period estimation $\approx$0.13 day \citep{2011ApJ...743...48W, 2018MNRAS.476.1224A,2020A&A...638A..20M}.
Assuming tidal locking (i.e., the orbital period equals the rotation period of the M star), the activity index and estimated rotation period locate TIC16320250 in the saturation region in the famous activity-rotation relation \citep{2011ApJ...743...48W, 2020ApJ...902..114W}.
Figure \ref{activity.fig} plots the comparison between these different indices (i.e., $R_{X}$, $R_{HK}$, $R_{NUV}$, and $R_{FUV}$), indicating the excess in UV luminosities is mainly contributed by the M dwarf and the temperature of the white dwarf companion is much less than 11250 $K$ (Section \ref{nat.sec}).
In summary, the X-ray, Ca H\&K, and UV emissions show that TIC 16320250 is one of the most magnetically active stars.

\begin{figure*}[htbp!]                                                            
\center 
\includegraphics[width=0.9\textwidth]{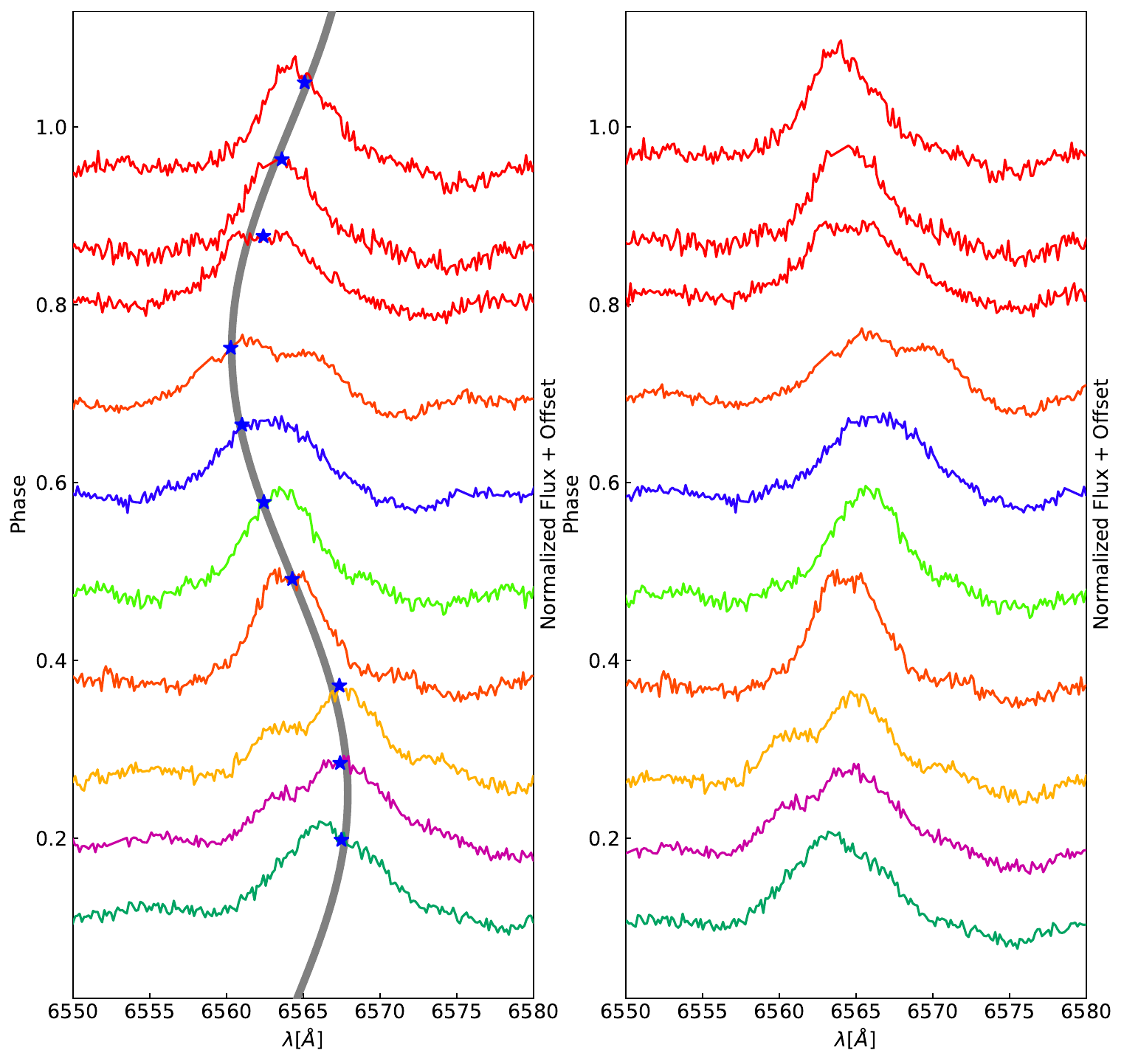}
\caption{Left panel: phased observed H$\alpha$ emission-line profiles. The blue stars are radial velocities of each spectrum, and the gray line shows the theoretical radial velocity curve from the {\it The Joker}. Right panel: phased RV-corrected H$\alpha$ emission-line profiles.}
\label{Ha.fig}
\end{figure*}

In addition, Figure \ref{Ha.fig} shows the broad and multi-component profiles of H$\alpha$ emission lines of this system. 
Besides the H$\alpha$ emission from the M star, also due to stellar activity, there is another component that may be from the accretion disk around the white dwarf. 
We tried a double-gaussian and triple-gaussian fitting to H$\alpha$ emission lines. However, the emission line from the M star is very broad and varies from exposure to exposure, making it difficult to accurately measure the velocity of the disk.

\section{Summary}
\label{summary.sec}

TIC16320250 is a semi-detached binary system with an orbital period of 0.25567 day.
The visible star is a lobe-filling M star ($f \sim$ 1) with a mass of 0.56 $M_{\odot}$.
The size of the Roche lobe and the profile of H$\alpha$ emission lines imply the potential for mass transfer and the presence of a disk around the unseen star.
Through radial velocity and light curve fitting, we confirm the unseen star to be a white dwarf with a mass of 0.67 $M_{\odot}$.

One intriguing aspect of this system is the repeated pattern observed in the light curves, which suggests a possible stellar cycle.
As previously mentioned, compared with single stars, the orbital motion of the binary can be utilized as an anchor to accurately position the stellar spot and verify its stability over a long timescale.
By using a cross-correlation analysis, we identify a ten-year period of the repeated pattern.
The variability of the light curves is caused by the motion and/or appearance (and disappearance) of stellar spots, particularly polar spots, as revealed by our light curve fitting.
Furthermore, the M star is confirmed to be a magnetically active star based on its X-ray, Ca H\&K, and UV emissions.
This finding supports the variable behavior of the spots, which are also correlated to stellar magnetic fields.
Recently, \citet{2023arXiv230711146R} identified long-term light curve variations, evidencing for spot modulations or star cycles, in two binaries including a K dwarf and a white dwarf.
More directly, Doppler imaging of the binary V471 Tau, consisting of a K2 dwarf and a white dwarf, shows a polar spot and a low-latitude spot on the surface of the K dwarf \citep{1995AJ....110.1364R}, similar to TIC16320250.
Therefore, light curve modeling, capable of revealing the evolution of stellar spots, can be a valuable tool for tracing the stellar activity cycle.
This method holds promise for future applications in the SiTian Project  \citep{2021AnABC..93..628L} to explore stellar cycles in a number of stars, especially those in binary systems.

In such a close binary, however, the differential rotation on the surface of the M star might be suppressed.
Thus the mechanism for the motion of the polar spots becomes a matter of consideration.
We proposed that the strong tidal force from the white dwarf companion could serve as a contributing factor.
The tidal force could lead to a 1:1 resonant excitation of the oscillation of the $\alpha$-effect, which is capable of exciting the underlying dynamo \citep{2016SoPh..291.2197S,2019SoPh..294...60S,2023arXiv230105452K}, though this is still debated \citep{2005SoPh..229..175D,2022SoPh..297..107N,2023A&A...671A..87W}.
In binary systems, the tidal interactions between binary components are much larger than tidal effects from planets and are expected to induce large-scale 3D shear and/or helical flows in stellar interiors that can significantly perturb the stellar dynamo.
On the other hand, turbulent Ohmic dissipation of magnetic flux may play an important role in stellar dynamo \citep{2022MNRAS.513.5474W}.
They could lead to the formation of clusters of flux tube eruptions at preferred longitudes, which result from the cumulative and resonant character of the action of tidal effects on rising flux tubes \citep{2003A&A...405..291H,2003A&A...405..303H}.
These processes may finally affect the behavior of stellar spots and the timescale of stellar cycles.

\begin{acknowledgements}

We thank the anonymous referee for helpful comments and suggestions that have improved the paper. 
The Guoshoujing Telescope (the Large Sky Area Multi-Object Fiber Spectroscopic Telescope LAMOST) is a National Major Scientific Project built by the Chinese Academy of Sciences. Funding for the project has been provided by the National Development and Reform Commission. LAMOST is operated and managed by the National Astronomical Observatories, Chinese Academy of Sciences. 
This work uses data obtained through the Telescope Access Program (TAP), which has been funded by the TAP member institutes. 
This work presents results from the European Space Agency (ESA) space mission {\it Gaia}. {\it Gaia} data are being processed by the {\it Gaia} Data Processing and Analysis Consortium (DPAC). Funding for the DPAC is provided by national institutions, in particular the institutions participating in the {\it Gaia} Multilateral Agreement (MLA). The {\it Gaia} mission website is https://www.cosmos.esa.int/gaia. The {\it Gaia} archive website is https://archives.esac.esa.int/gaia. We acknowledge use of the VizieR catalog access tool, operated at CDS, Strasbourg, France, and of Astropy, a community-developed core Python package for Astronomy (Astropy Collaboration, 2013). This research made use of Photutils (Bradley et al. 2020), an Astropy package for detection and photometry of astronomical sources. This work was supported by National Science Foundation of China (NSFC) under grant Nos. 11988101/11933004/11833002/12090042/12103047/
12273057, National Key Research and Development Program of China (NKRDPC) under grant Nos. 2019YFA0405504 and 2019YFA0405000, and Strategic Priority Program of the Chinese Academy of Sciences under grant No. XDB41000000. 
We acknowledge the Science Research Grants from the China Manned Space Project with No. CMS-CSST-2021-A08. 
The present study is also financially supported by Yunnan Fundamental Research Projects (grant Nos. 202201AT070186 and 202305AS350009), and International Centre of Supernovae, Yunnan Key Laboratory (No. 202302AN360001). S.W. acknowledges support from the Youth Innovation Promotion Association of the CAS (IDs 2019057).

\end{acknowledgements}

\bibliography{main.bib} 

\clearpage
\appendix
\renewcommand*\thetable{\Alph{section}.\arabic{table}}
\renewcommand*\thefigure{\Alph{section}\arabic{figure}}

\section{RV measurements of TIC 16320250}
\label{rvdata_appendix.sec}

\setcounter{table}{0}

\begin{longtable}{cccc|cccc}
\caption{Barycentric-corrected RV values of TIC 16320250. \label{rvdata.tab}}
\\\hline\noalign{\smallskip}
BMJD & RV & Uncertainty & Instrument & BMJD & RV & Uncertainty & Instrument \\
(day) & (km/s) & (km/s)  & & (day) & (km/s) & (km/s) & \\
\hline\noalign{\smallskip}
58182.86136 & -98.81 & 3.26 & LAMOST/MRS & 59681.73791 & 15.29 & 3.90 & 2.16 m/BFOSC \\
58182.87733 & -40.28 & 5.54 & LAMOST/MRS & 59681.75886 & -41.44 & 3.80 & 2.16 m/BFOSC \\
58182.89331 & 19.26 & 3.35 & LAMOST/MRS & 59681.77980 & -106.88 & 4.60 & 2.16 m/BFOSC \\
58186.84557 & -10.76 & 3.64 & LAMOST/MRS & 59681.80075 & -180.40 & 3.95 & 2.16 m/BFOSC \\
58186.86155 & -90.31 & 5.25 & LAMOST/MRS & 59681.82168 & -203.11 & 4.12 & 2.16 m/BFOSC \\
58186.87821 & -164.35 & 3.35 & LAMOST/MRS & 59681.84261 & -167.53 & 3.02 & 2.16 m/BFOSC \\
58186.89419 & -194.37 & 3.54 & LAMOST/MRS & 59681.86356 & -58.37 & 4.88 & 2.16 m/BFOSC \\
59294.83498 & 135.33 & 4.71 & LAMOST/MRS & 59692.62962 & 16.50 & 3.75 & 2.16 m/BFOSC \\
59294.85096 & 129.33 & 3.26 & LAMOST/MRS & 59692.65057 & 95.69 & 3.25 & 2.16 m/BFOSC \\
59294.86762 & 129.83 & 3.07 & LAMOST/MRS & 59692.67152 & 141.77 & 3.50 & 2.16 m/BFOSC \\
57090.82541 & -162.35 & 3.92 & LAMOST/LRS & 59692.69247 & 142.71 & 3.30 & 2.16 m/BFOSC \\
57090.83444 & -182.86 & 3.88 & LAMOST/LRS & 59692.71340 & 103.45 & 2.77 & 2.16 m/BFOSC \\
57090.84347 & -192.87 & 3.84 & LAMOST/LRS & 59692.73449 & 10.45 & 3.80 & 2.16 m/BFOSC \\
58901.87328 & 128.83 & 4.65 & LAMOST/LRS & 59692.75544 & -70.79 & 3.10 & 2.16 m/BFOSC \\
58901.88231 & 141.84 & 4.36 & LAMOST/LRS & 59692.77639 & -161.38 & 3.00 & 2.16 m/BFOSC \\
58901.89203 & 142.84 & 4.53 & LAMOST/LRS & 59692.79734 & -209.49 & 2.83 & 2.16 m/BFOSC \\
58901.90106 & 145.84 & 4.32 & LAMOST/LRS & 59692.84038 & -138.94 & 3.22 & 2.16 m/BFOSC \\
58901.91009 & 133.33 & 4.24 & LAMOST/LRS &  &  &  & \\
\noalign{\smallskip}\hline
\end{longtable}

\section{PHOEBE fitting to light curves.}
\label{pfit.sec}

\setcounter{figure}{0}

\begin{figure*}[htbp!]  
\center 
\includegraphics[width=1\textwidth]{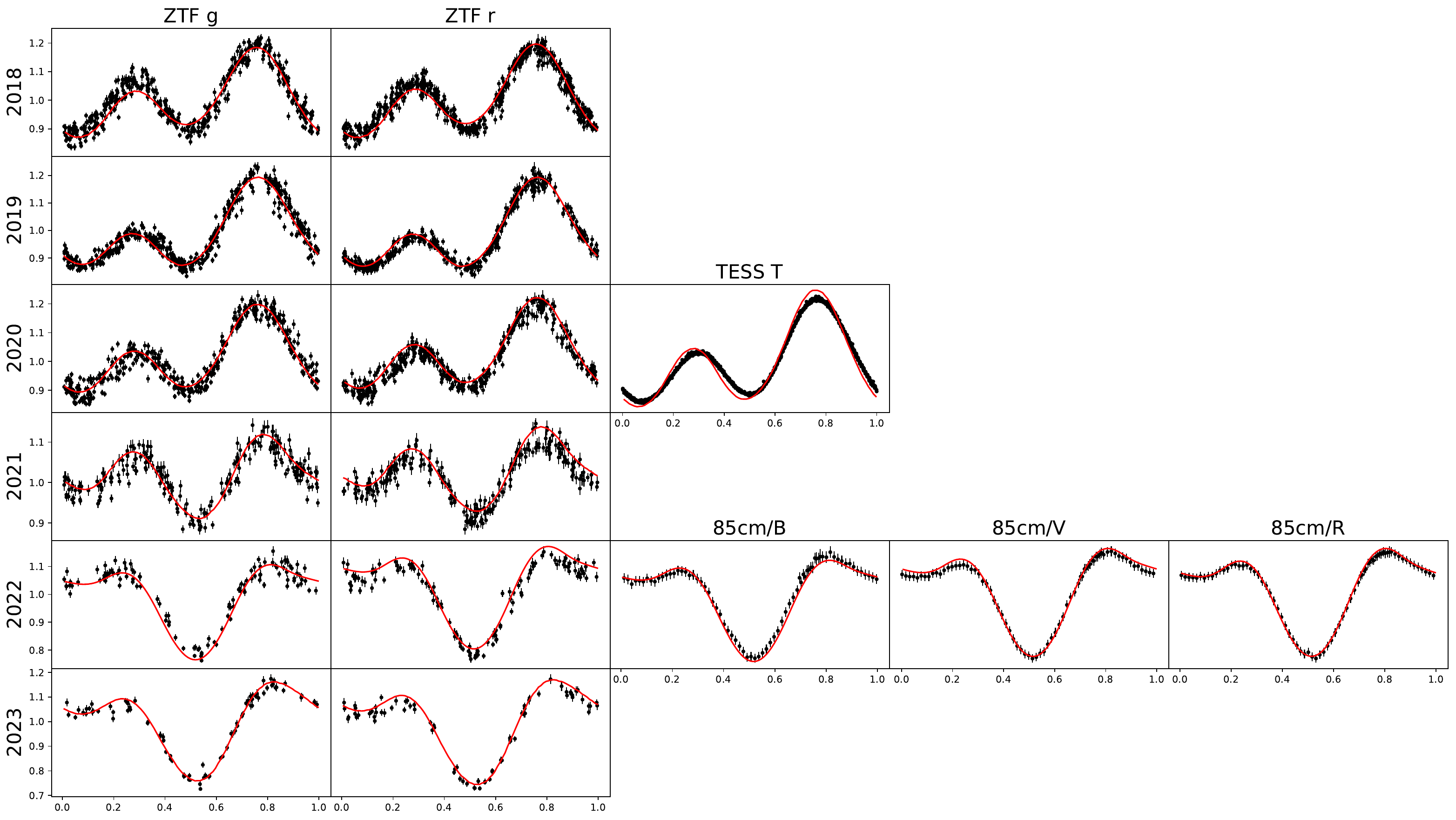}
\caption{PHOEBE fitting (model A; red lines) to the normalized light curves from ZTF $g$ band, ZTF $r$ band, TESS, and 85 cm $B$/$V$/$R$ band.}
\label{fit_lcs.fig}
\end{figure*}

\begin{figure*}[htbp!]  
\center 
\includegraphics[width=1\textwidth]{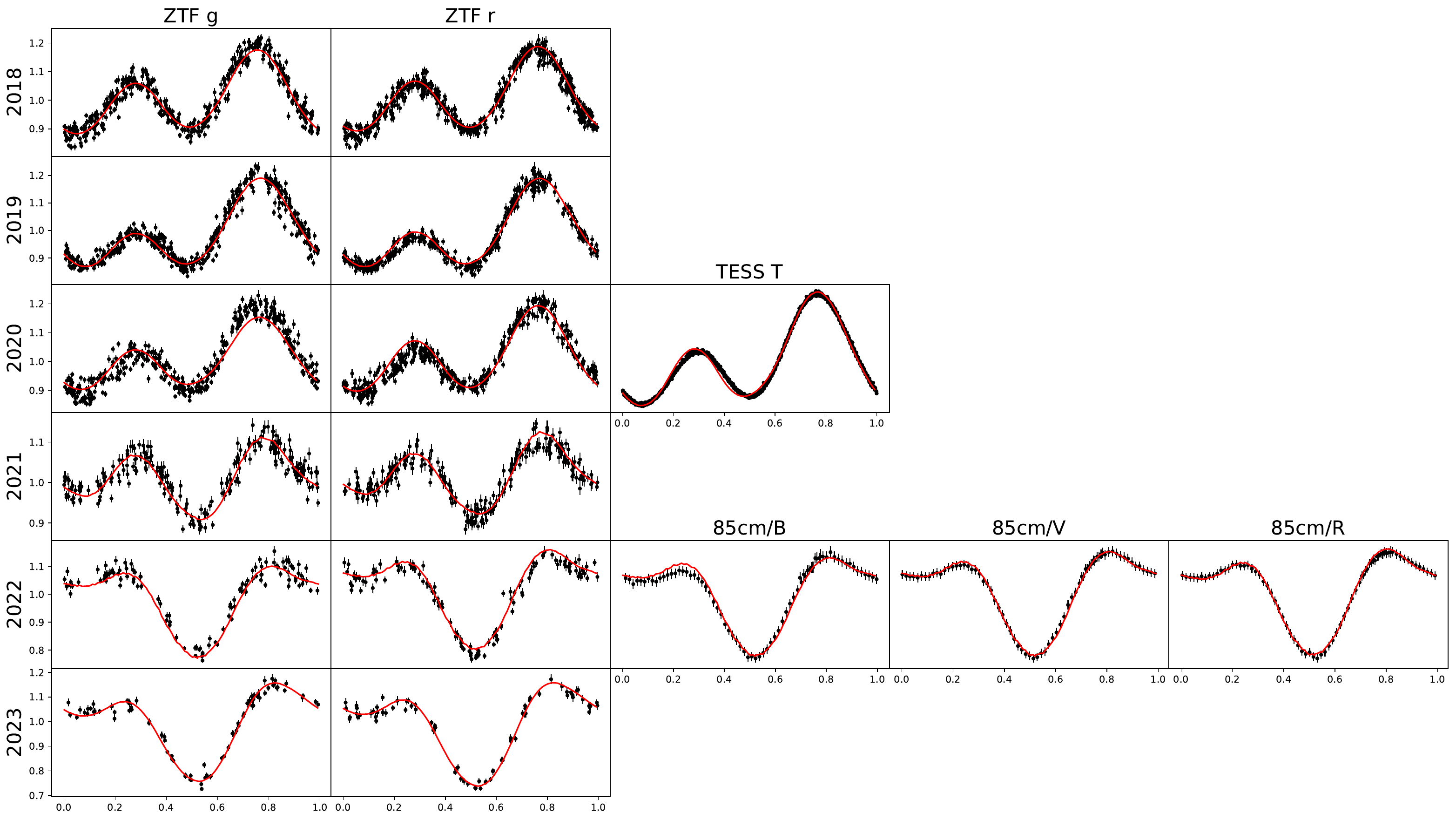}
\caption{PHOEBE fitting (model B; red lines) to the normalized light curves from ZTF $g$ band, ZTF $r$ band, TESS, and 85 cm $B$/$V$/$R$ band.}
\label{fit_lcs_2spots.fig}
\end{figure*}

\end{document}